\documentclass[prl,superscriptaddress,twocolumn]{revtex4}
\usepackage{epsfig}
\usepackage{latexsym}
\usepackage{amsmath}
\begin {document}
\title {Travelling Salesman Problem with a Center}
\author{Adam Lipowski}
\affiliation{Faculty of Physics, Adam Mickiewicz University, 61-614
Pozna\'{n}, Poland}
\author{Dorota Lipowska}
\affiliation{Institute of Linguistics, Adam Mickiewicz University,
60-371 Pozna\'{n}, Poland}
\begin {abstract}
We study a travelling salesman problem where the path is optimized
with a cost function that includes its length $L$ as well as a
certain measure $C$ of its distance from the geometrical center of
the graph. Using simulated annealing (SA) we show that such a
problem has a transition point that separates two phases differing
in the scaling behaviour of $L$ and $C$, in efficiency of SA, and
in the shape of minimal paths.
\end{abstract}
\maketitle
In the travelling salesman problem (TSP), which is perhaps the
most famous combinatorial optimization problem, one has to find
the shortest path that joins a given set of $N$ points. In
addition to pure academic interest TSP appears naturally in some
transportation applications or production and testing of
integrated circuits~\cite{JOHNSON}. This easy to formulate problem
is however very difficult to solve. Actually TSP is known to be
NP-complete~\cite{GAREY} and it is believed that there is no
algorithm that can find a solution in time increasing as a finite
order polynomial in $N$. Although advanced algorithms can find
exact solutions of TSP for quite a large values of $N$, these
algorithms are usually dedicated only to this particular task.
Consequently, they are not suitable for more general versions of
TSP nor for some other numerically diffcult optimization problems.
It is thus desirable to develop approximate but more versatile
methods such as genetic programming~\cite{KOZA}, simulated
annealing (SA)~\cite{KIRK} or extremal optimization~\cite{PERCUS},
and for such a purpose TSP appears to be an excellent testing
ground.

Another NP-complete problem that is of considerable interest is
the so-called satisfiability problem where one examines conditions
needed to satisfy certain Boolean formulae~\cite{HAYES}. Recently,
it was shown that there is a phase transition in this class of
problems that separates two regimes: easy and hard to satisfy. It
turns out that such problems are most difficult to examine right
at the transition point and computational complexity decreases
when one moves away of the transition point~\cite{HOGG}. A related
phase transition was found in a certain version of TSP where
distances between points are randomly drawn integer numbers
$\{1,2,..k\}$. In such a case the time needed to find the minimal
solution using branch-and-bound algorithm dramatically increases
when $r$ exceeds a certain threshold value~\cite{ZHANG}. Methods
used to establish these results very often originate from
statistical mechanics providing thus an interesting
multidisciplinary bridge~\cite{MONASSON}.Although some other
examples of connections between statistical mechanics and
computational complexity were already examined~\cite{MERTENS},
further explorations of this subject would be very desirable.

In the present paper we examine a certain version of TSP where one
requires the minimization of $E=L+rC$, where $L$ is the total
length of a path, $C$ is the sum of distances of middle points of
links to a geometrical center of the graph, and $r$ is a control
parameter of the model (of course, $r=0$ corresponds to the
original TSP). Using simulated annealing we show that upon
increasing $r$ the model undergoes a phase transition from the
$L$(-dominated) phase (small $r$) to the $C$(-dominated) phase
(large $r$). In the $L$-phase we find $L\sim N^{1/2}$ and $C\sim
N$, while in the $C$-phase a reverse scaling holds with $L\sim N$
and $C\sim N^{1/2}$. Moreover, minimal paths in these two phases
have qualitatively different shape. What is also interesting is
the change of efficiency of SA. The $r$-dependent optimization
problem that we study does not seem to be easier than the $r=0$
case, and we expect that for any $r$ our problem is also
NP-complete. Nevertheless, in the $C$-phase SA has a much better
efficiency than in the $L$-phase at least with respect to finding
nearly-minimal paths. Such an effect might be related with a
change of the energy landscape of the problem, but further studies
would be needed to verify such a claim.

To define our problem, let us consider $N$ points distributed in a
unit square of a Euclidean plane. Denoting coordinates of $i$-th
point as $(x_i, y_i)$, we have
{\setlength\arraycolsep{0pt}
\begin{eqnarray}
L & = &\sum_{k=1}^{N}\sqrt{(x_{i_k}-x_{i_{k+1}})^2+(y_{i_k}-y_{i_{k+1}})^2},\\
C & = &
\frac{1}{2}\sum_{i=1}^{N}\sqrt{(x_{i_k}+x_{i_{k+1}}-1)^2+(y_{i_k}+y_{i_{k+1}}-1)^2},
\label{eq_lc}
\end{eqnarray}}
where we assume that in a given path, points appear in the order
$i_1,i_2,\ldots,i_N,i_1$ and $i_{N+1}=i_1$. Next, we introduce the
cost function $E=L+rC$. We cannot provide an immediate application
of such a problem, but one can imagine that in some transportation
tasks staying during the tour close to (or far from ) the center
might be of some importance. Such  an additional condition might
be particularly  relevant in war areas where staying near a
military base is important for safety reason. Another application
might be a "just-in-time vehicle routing problem", in which truck
drivers might be called to the center depot intermediately in
order to get further goods that have to be delivered to the
customers. It is easy to realize that minimization of $L$ usually
does not minimize $C$ and competition of these two terms might
lead to some interesting effects.

To find a path that minimizes $E$ we use a standard simulated
annealing technique. For a given configuration, we produce a trial
configuration by exchanging randomly positions of two points. Such
a move is accepted with a probability min$(1,e^{-\delta E/T})$
where $\delta E$ is the energy difference between a trial and
initial configuration, and $T$ is a fictitious temperature. During
simulations, the temperature $T$ is reduced to zero linearly in
time $t$ according to $T(t)=1-ct$ where $c$ is the cooling rate
and a unit of time is defined as an attempt to make $N$ moves. For
a small cooling rate $c$ such a method finds a path of a low (and
for small $N$ perhaps even lowest) value of $E$. To check that the
properties of our model are not the consequence of the numerical
method, we did additional runs using the so-called Lin-2-Opt moves
that cut a configuration in two pieces, turn the direction of one
piece around, and reconnect~\cite{LIN}. Such moves are known to be
more effective than exchanges of two points~\cite{STADLER}. Both
methods, however, yields essentially the same behaviour of our
model with respect to e.g., location of the transition point or
$N$-dependence of $L$ and $C$. Unless specified otherwise,
described below results are obtained with the use of the exchange
algorithm.

\begin{figure}
\centerline{ \epsfxsize=9cm \epsfbox{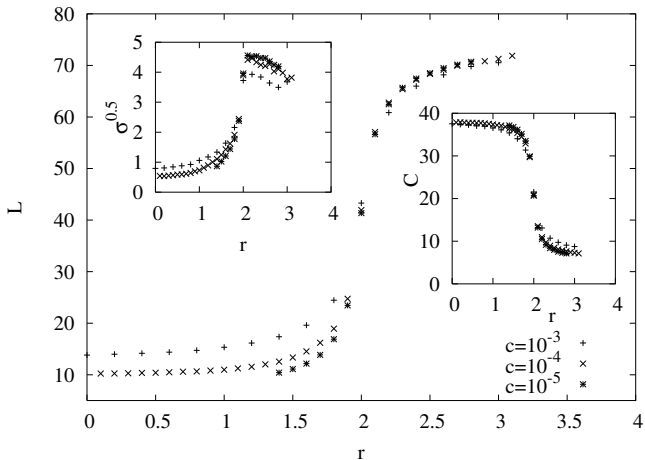} }
\caption{The length $L$ of the minimal path as a function of $r$
for $N=100$. The left inset shows square root of its variance
$\sqrt{\sigma}$ as a function of $r$. The right inset shows the
total distance from the center $C$ as a function of $r$.}
\label{variance}
\end{figure}

In Fig.~\ref{variance} we show the length $L$ of a minimal path as
a function of $r$ averaged over 1000 distributions of $N=100$
points. One can see that $L$ is a slowly increasing function of
$r$ except around $r=2.0$ $L$ where the increase is more abrupt.
Around the same value of $r$ there is the maximum of the variance
of $L$ and the abrupt change of $C$. These results suggest that a
certain phase transition takes place around $r=2$.

\begin{figure}
\centerline{ \epsfxsize=9cm \epsfbox{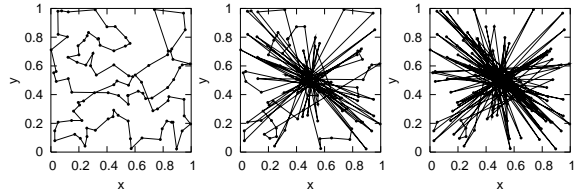} }
\caption{Typical shapes of nearly minimal paths for (from left to
right) $r=0.0$, 2.0 and 2.5. Calculations were made for $N=100$
and $c=10^{-6}$.} \label{config}
\end{figure}

A visual indication of a qualitative change around $r\sim 2.0$ is
shown in Fig.~\ref{config}. For large $r$ ($C$-dominated phase),
paths that minimize $E$ have large length $L$ but many
intersections around the center ($1/2,1/2$) yield a small value of
$C$. For small $r$ ($L$-dominated phase) typical minimal paths have a much different shape.
They have only few intersections and are more or less uniformly distributed in a
unit square.

\begin{figure}
\centerline{ \epsfxsize=9cm \epsfbox{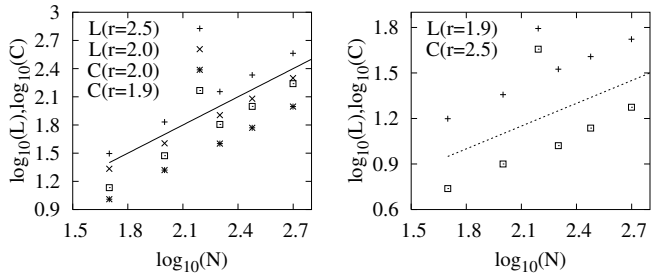} }
\caption{Size dependence of $L$ and $C$ for $r=1.9$, 2.0 and 2.5.
Calculations made for $c=10^{-4},\ 10^{-5}$ and $10^{-6}$ were extrapolated to $c=0$ and average was made over
1000 independent samples. Continuous and dotted lines have slope
1.0 and 0.5, respectively.} \label{size}
\end{figure}

That in our problem there are two phases separated by a transition
point around $\sim 2.0$ is also seen in Fig.~\ref{size} which
presents the $N$-dependence of $L$ and $C$. These results show
that in the $L$-phase ($r=1.9$) $L\sim N^{1/2}$ and $C\sim N$.
Consequently, the cost function $E$ is dominated by the distance
$C$. Let us notice that the scaling $L\sim N^{1/2}$ was already
proven for the original TSP ($r=0$)~\cite{BHH}. As for the second
relation ($C\sim N$), there is a simple argument that  justifies
it. Indeed, since in the $L$-phase a minimal path is relatively
uniformly distributed in the unit square one can assume that
central points of links are also almost uniformly distributed. It
means that there is a positive (and $N$-independent in the limit
$N\rightarrow\infty$) average distance of a central point from
(1/2,1/2) and then $C\sim N$ easily follows~\cite{COMM1}.

\begin{figure}
\centerline{ \epsfxsize=9cm \epsfbox{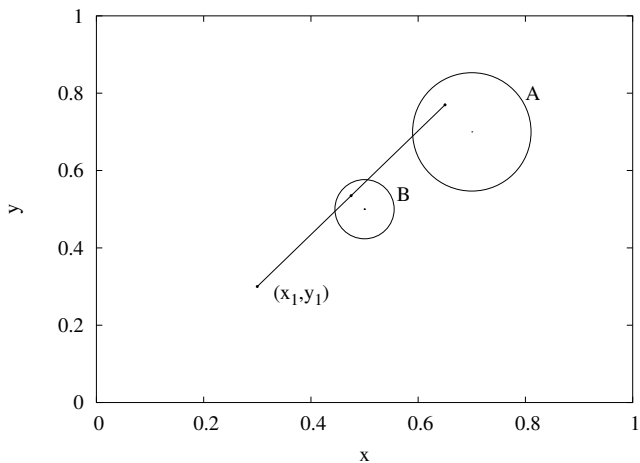} }
\caption{A link with a point inside A circle has a center inside B
circle with a twice smaller radius.} \label{rysunek}
\end{figure}

\begin{figure}
\centerline{ \epsfxsize=9cm \epsfbox{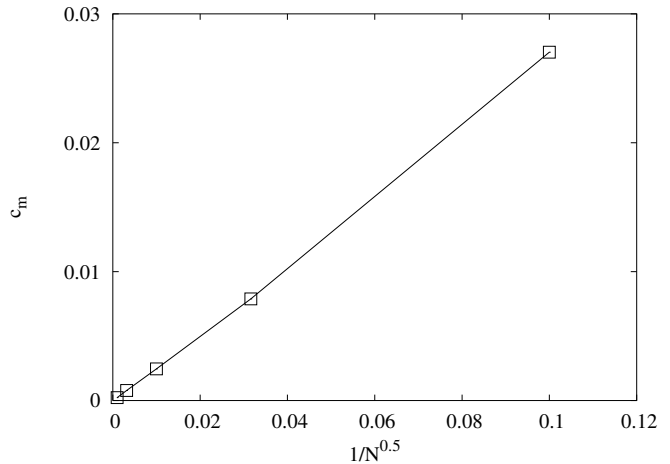} }
\caption{The minimal distance $c_m$ from a center of a link to
(1/2,1/2) as a function of $1/\sqrt{N}$. Links are made between a
randomly chosen point ($x_1,y_1$) and $N-1$ other randomly chosen
points. Presented results are averages over 100 choices of
($x_1,y_1$).} \label{rysunek1}
\end{figure}

In the $C$-phase we find $L\sim N$ and $C\sim N^{1/2}$. Such
relations can be deduced from the structure of a typical minimal
path as seen in Fig.~\ref{config}. In the $C$-phase points that
are approximately opposite with respect to (1/2,1/2) are pairwise
linked. It means that there is a finite ($N$-independent) length
of such a link and thus $L\sim N$ follows. The relation $C\sim
\sqrt{N}$ is more subtle and we can justify it only approximately.
In the following we argue that for a minimal path a typical
distance $\bar{c}$ of a center of a link to (1/2,1/2) scales as
$1/\sqrt{N}$ from which the relation $C\sim \sqrt{N}$ easily
follows. To show that $\bar{c}\sim 1/\sqrt{N}$ let us consider
e.g., a point $(x_1,y_1)$ and ask what is a minimum $c_m$ of
distances of a center of a link between $(x_1,y_1)$ and any of the
remaining $N-1$ points. It means that we have to find a minimum of
$\frac{1}{2}\sqrt{(x_1+x_i-1)^2+(y_1+y_i-1)^2}$ where
$i=2,3\ldots,N$. Let us notice that in a (two-dimensional) unit
square $N-1$ points set a characteristic distance between points
as $1/\sqrt{N}$. Thus one can expect that in a circle of a radius
$1/\sqrt{N}$ that is opposite (with respect to (1/2,1/2)) to the
point ($x_1,y_1$) there is approximately one of these $N-1$ points
(see Fig.~\ref{rysunek}). A link with such a point has a center
that is inside a circle around (1/2,1/2) and has a twice smaller
radius (but still it is of the order of $1/\sqrt{N}$).
Consequently, the smallest distance $c_m$ of a center of link to
the point ($1/2,1/2$) should scale as $1/\sqrt{N}$ and this
argument is supported by simple numerical calculations that we
present in Fig.~\ref{rysunek1}. We expect that in the minimal path
$\bar{c}$ also scales as $c_m$ i.e., as $1/\sqrt{N}$ and thus the
relation $C\sim \sqrt{N}$ follows.
Let us notice that scaling of $L$ and $C$ in the $C$-phase is
analogous to the scaling in the $L$-phase but with the role of $L$
and $C$ interchanged. We do not know whether this relation is
accidental or indicates a deeper relation (e.g., certain duality)
between these two phases.

As it might be expected, the scaling of $L$ and $C$ at the
transition point $r=2.0$ is different than in each of the phases.
Figure~\ref{size} shows that in this case both quantities increase
linearly with $N$.

\begin{figure}
\centerline{ \epsfxsize=9cm \epsfbox{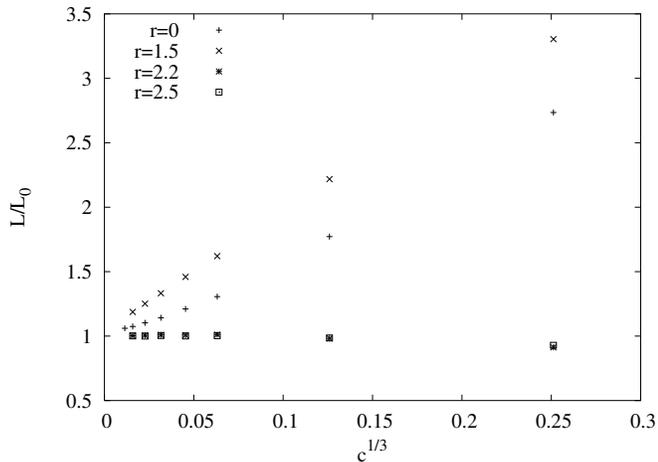} }
\caption{The average normalized length $L/L_0$ as a function of
cooling rate $c$, where $L_0$ is the extrapolated value of $L$ at
the zero cooling rate ($N=100$).}
\label{coola}
\end{figure}
\begin{figure}
\centerline{ \epsfxsize=9cm \epsfbox{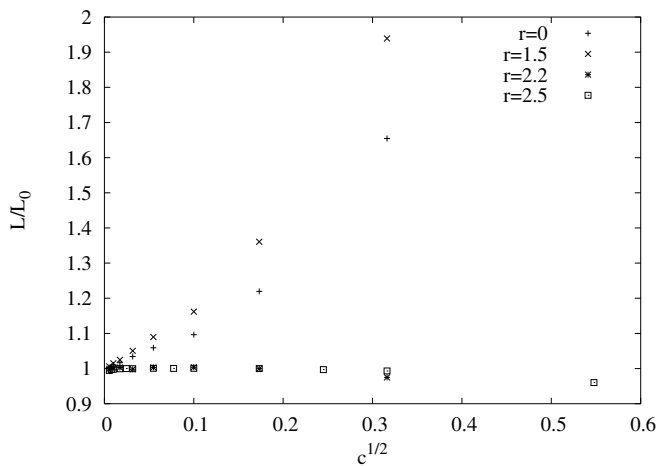} }
\caption{The average normalized length $L/L_0$ as a function of
exponential cooling rate $c$, where $L_0$ is the extrapolated
value of $L$ at the zero cooling rate ($N=200$). Simulations were
made using Lin-2-Opt moves. Let us notice that $L$ converges to
$L_0$ linearly in $c^{1/2}$. For the algorithm with exchange moves
(Fig.~\ref{coola}) convergence is slower and linear in $c^{1/3}$.}
\label{lin2}
\end{figure}

Finally, we compare the efficiency of the simulated annealing in
each phase. Our data (Fig.~\ref{coola}) show that in the $C$-phase
SA is much more efficient and already with a relatively fast
cooling a (nearly) minimal path is found. In the $L$=phase
corrections due to the nonzero cooling-rate are much more
important (our simulations show that these corrections in the
$L$-phase decay approximately as $c^{1/3}$). For $r=0$ our data
extrapolated to the limit of zero cooling rate give for $N=100$
the average length of a minimal path as $L_0=7.8$, which is in a
good agreement with other results quoted in the
literature~\cite{LEE}. For comparison, we show also the results
obtained using the Lin-2-Opt algorithm~\cite{LIN}. This time
simulations were made for $N=200$ and the exponential cooling
schedule was used $T(t)=T_0(1-c)^t$, whose effectiveness was
already demonstrated~\cite{SCHRIMPF}. Results of these
simulations, shown in Fig.~\ref{lin2}, confirm the change of
efficiency of SA at the transition point of our model.

A main difficulty with finding a minimal path in TSP is due to the
complexity of the cost function and more specifically due to its
many local minima. Some of these minima, being quite far from the
global minimum, might be at the same time quite deep and the
searching algorithm might get stuck in one of them. The landscape
of the cost function in our problem is also very complex and has
many minima. However, a substantial increase in efficiency of SA
in the $C$-phase indicates a certain change in this landscape. One
possibility is that in the $C$-phase local minima are located
mainly in the vicinity of the global minimum. As a result, a
solution found using SA, which is almost always only a nearly
optimal path, is a good approximation of the optimal solution, In
our opinion, such a transition for a model that most likely is
NP-complete for any $r$ is new and it would be desirable to have
its better understanding. Another type of a transition in TSP
appears in the time-dependent version of this
problem~\cite{BENTNER}.

In conclusion, we studied a version of the travelling salesman
problem where the cost function includes both the length of a path
as well as its distance from a center. This problem, depending on
the control parameter, turns out to have two phases with different
kinds of solutions. As indicated by a drastic change in the
efficiency of simulated annealing method, a phase transition in
this problem should be accompanied by important changes in the
complex landscape of the cost function. Further studies would be
needed to clarify more detailed properties of this problem.

Acknowledgments: The research grant 1 P03B 014 27 from KBN is
gratefully acknowledged.
 Numerical calculations were
partially performed on ' Open Mosix Cluster' built and
administrated by Dr.~L.~D\c{e}bski.


\begin{thebibliography}{}
\bibitem{COMM1} Assuming that distribution of central points of links is
uniform in the unit square we obtain that their average distance
from (1/2,1/2) equals $\bar{c}=\int_0^1 \int_0^1
\sqrt{(x-0.5)^2+(y-0.5)^2}dxdy\sim 0.384$. For $r=0$ and $N=100$ our
numerical calculations give $C\sim 37.3$ which is in a relatively
good agreement with the approximation $C=\bar{c}N$.
(As might be expected, it is slightly smaller.)
\end{thebibliography}
\end {document}